\documentclass{epl}

\title{Incomplete ordering of the voter model on small-world networks}
\shorttitle{Incomplete ordering of the voter model...}
\author{Claudio Castellano\inst{1,2}
\thanks{E-mail: \email{castella@pil.phys.uniroma1.it}},
Daniele Vilone\inst{1}
\and Alessandro Vespignani\inst{3}}
\shortauthor{C. Castellano \etal}
\institute{
 \inst{1} Dipartimento di Fisica, Universit\`a di Roma ``La Sapienza'', 
P. le A. Moro 2, I-00185 Roma, Italy \\
 \inst{2} INFM, Unit\`a di Roma 1, P. le A. Moro 2, I-00185 Roma, Italy \\
 \inst{3} Laboratoire di Physique Th\'eorique (UMR du CNRS 8627) 
B\^atiment 210, Universit\'e de Paris-Sud, 91405 ORSAY Cedex, France \\
}


\begin{document}
\maketitle

\begin{abstract}
We investigate how the topology of small-world networks affects the
dynamics of the voter model for opinion formation.
We show that, contrary to what occurs on regular topologies with local
interactions, the voter model on small-world networks
does not display the emergence of complete order in the thermodynamic limit.
The system settles in a stationary state with coexisting opinions
whose lifetime diverges with the system size.
Hence the nontrivial connectivity pattern leads to the counterintuitive
conclusion that long-range connections inhibit the ordering process.
However, for networks of finite size, for which full uniformity is reached,
the ordering process takes a time shorter than on a regular lattice of the
same size.
\end{abstract}

In the last decade, social sciences have started to deal with
large scale modeling of a variety of spreading and ordering phenomena
that involve cooperative behavior~\cite{Axelrod97}.
In this context, classical models developed in statistical physics 
to study the onset of order in matter~\cite{Bray94} have turned out to be 
useful for the investigation of the principles at the basis of social ordering.
For instance, the Ising and voter models and their variations are
prototypical models for a wide class of social interaction
phenomena~\cite{Axtell96, Castellano00, Lopez01, Holyst01, Stauffer02,
Liggett85,Vazquez02}.
The voter model is possibly the minimal model for opinion spreading
and the study of the onset of consensus.
It is usually defined on a regular lattice of dimension $d$.
Each site is characterized by a discrete variable $s$ that may
assume two values ($s = \pm 1$) representing two opposite opinions; for
instance the electoral choice in favor of two different candidates.
Starting from a disordered initial configuration, the model follows a simple
dynamical evolution in which at each time step one site is selected at
random and made equal to one of its nearest neighbors (chosen at random
on its turn). This dynamics mimics the homogenization of opinions
through the confrontation of peers and leads
to the formation and coarsening of ordered regions where individuals
share the same opinion.
In $d=1$ and $2$ the model eventually converges to an ordered state
with all variables having the same value~\cite{note1}.
This state is {\em absorbing} since the system cannot escape from it
once it is reached~\cite{Marrobook}.

While the use of regular lattices to model the interaction between
elementary objects is well justified for most physical situations, 
such an assumption is not obvious in the context of social sciences. 
Many social systems indeed show interaction patterns that find 
a better characterization as complex networks with distinctive 
connectivity properties~\cite{Strogatz01,Barabasi02}. 
In a graph representation, where nodes identify the individuals and
links their direct interactions, many social and natural networks exhibit
peculiar topological properties related to the presence of highly connected
individuals and long range connections. Among these features, the
most well documented is the small diameter of social
networks, i. e. each individual can reach any other one
passing through a very small number of intermediate nodes.
In addition, social interactions favor the connection between
common acquaintances leading to the presence of high clustering among 
nodes. This is quantitatively expressed by a high probability that if
two nodes share a neighbor they are directly connected on their turn.
This last property, along with the small diameter, define the so-called
small-world behavior~\cite{Watts98,Wattsbook}. 
The use of small-world like topologies in models of opinion spreading
is then a logical step in the direction of a more realistic
approach to the phenomenon. The interaction patterns involved in the
process of opinion formation are very likely similar to those of social 
networks such as the web of sexual contacts~\cite{Liljeros01} or
scientific collaborations~\cite{Redner98}.

In order to investigate how complex connectivity patterns 
might influence opinion formation, we  study the effect of the
small-world topology on the evolution of the voter model. 
The prototypical network possessing the small-world
character is the Watts-Strogatz (WS) model that has been extensively
studied in several contexts~\cite{Watts98,Wattsbook,Newman00}.
In particular, we have considered the WS network as defined in
Ref.~\cite{Watts98}.
Starting from a one-dimensional lattice of $N$ sites with periodic boundary
conditions and each node connected with $2 k$ nearest neighbors,
a stochastic rewiring is introduced. Nodes are visited one by one
sequentially and each of the $k$ links connecting the node to its nearest
neighbors in the clockwise sense is rewired with probability $p$
to a randomly chosen node.
As $p$ is increased, the WS network interpolates between a one-dimensional
lattice ($p=0$), with only geographical neighbors in contact, and a
random graph ($p=1$), where short and long range connections are equally
likely.
The small-world behavior (small diameter, high clustering)
is exhibited for values of the rewiring probability
$p$ such that $1/(k N) \ll p \ll 1$.
The transition between the one-dimensional topology and the small-world
one occurring for $p \approx 1/(k N)$ is governed by the value of $\xi$,
the average distance between nodes connected with shortcuts.
$\xi$ is the only nontrivial length in the network and can be
simply shown to scale as $1/(kp)$~\cite{Barthelemy99,Barrat00}.
If the network size $N$ is much smaller than $\xi$, that is $p \ll 1/(k N)$
the system does not have long-range connections and is a
one-dimensional lattice.
For $N \gg \xi$ many shortcuts are present and originate the small-world
behavior.

\begin{figure}
\onefigure[width=3.1in,angle=-90]{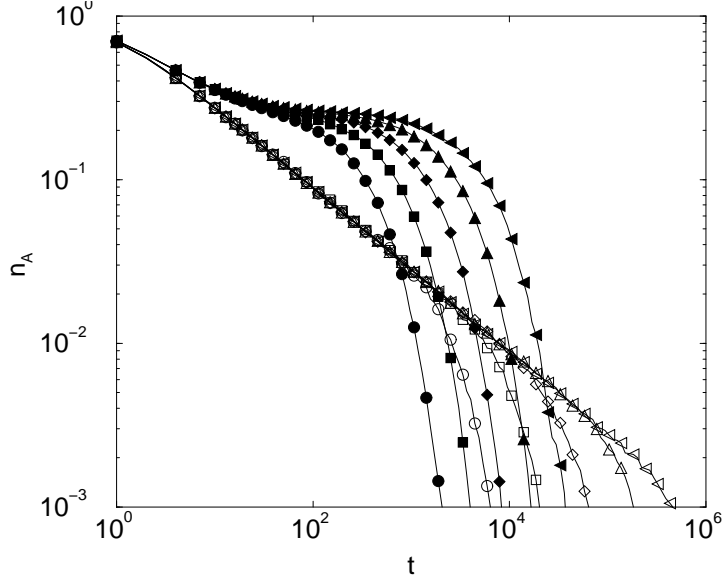}
\caption{Log-log plot of the fraction $n_A$ of active bonds between
nodes with different opinion. Values are averaged over 1000 runs.
Time is measured in Monte Carlo steps per site.
Empty symbols are for the one-dimensional case ($p=0$).
Filled symbols are for rewiring probability $p=0.05$. 
Data are for $N=200$ (circles), $N=400$ (squares), $N=800$ (diamonds),
$N=1600$ (triangles up) and $N=3200$ (triangles left).}
\label{Fig1}
\end{figure}

For the study of the ordering of the voter model,
the natural quantity of interest is the fraction $n_A$ 
of active bonds, i. e. the density of links connecting sites with
opposite values of $s$. These are the links where the dynamics takes place.
Such a quantity vanishes if the system orders entirely,
while it remains finite if domains coexist, and $1/n_A$ is a measure of
the average size of domains.
In the case $p=0$ (one-dimensional system), due to the diffusive motion
of interfaces between domains, the fraction of active
bonds decays with respect to time $t$  as
$n_A\sim t^{-1/2}$~\cite{Rednerbook}
up to a crossover time $t_0 \sim N^2$, after which $n_A$ exhibits an
exponential relaxation to the absorbing state $n_A=0$ (see Fig.~\ref{Fig1}).
Such a fast decay is the effect of the finite size of the system.
In the thermodynamic limit one recovers a pure power-law relaxation to
the absorbing state.

We consider now a Watts-Strogatz network with $k=2$, $p=0.05$ and values of
$N$ such that the system is safely inside the small-world regime
($N \geq 200$).
The behavior changes dramatically (Fig.~\ref{Fig1}):
After a transient, the plot of $n_A$ exhibits a plateau,
indicating that domains remain on average of constant size.
This regime is ended by an exponential approach to the absorbing state $n_A=0$.
The duration of the plateau grows with $N$,
but (as can be seen from Fig.~\ref{Fig1}) the time to reach complete
ordering is {\em smaller} for the WS network than for a regular lattice
with the same number of nodes.
This is in agreement with the naive expectation that long-range connections
should speed up the homogenization process.
However, this occurs in a highly nontrivial way:
During most of the evolution, $n_A$ is higher on the small-world network
than on a regular lattice, i. e. the small-world network is, for a long
time interval, more disordered, and orders rapidly only at the very end.
We have checked that this phenomenology is not an artifact of the 
rewiring procedure to build the Watts-Strogatz network: analogous
results are obtained when the small-world topology is produced by adding
random links to a one-dimensional lattice.

\begin{figure}
\onefigure[width=3.1in,angle=-90]{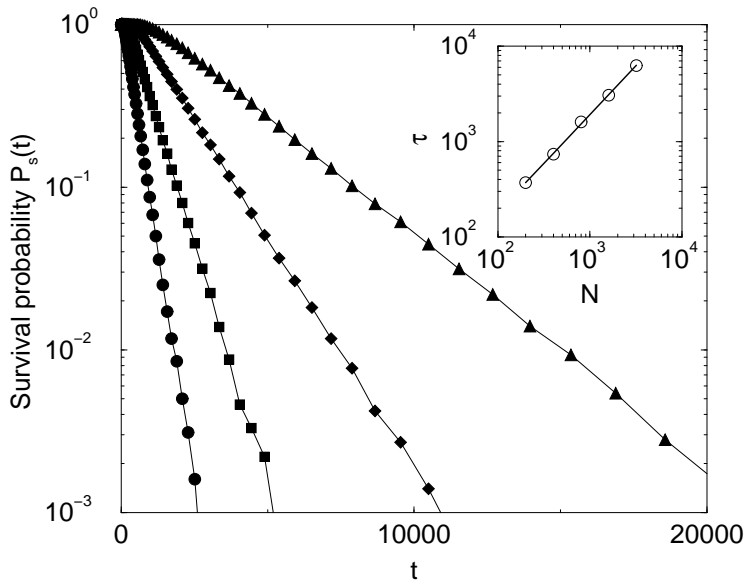}
\caption{Main: The survival probability $P_s(t)$ as a function of time
for a network with $p=0.05$.
Data are for $N=200$ (circles), $N=400$ (squares), $N=800$ (diamonds) 
and $N=1600$ (triangles) and are computed averaging over 1000 runs.
Inset: The average lifetime in the active state $\tau$ as a function of
the system size $N$. Circles are numerical values obtained by measuring
$\tau$ as the inverse slope of the curves shown in the main part of the
figure. The solid line is the best fit to the expression $\tau \propto
N^{\gamma}$ giving $\gamma=1.02 \pm 0.02$.}
\label{Fig2}
\end{figure}

The nature of the exponential approach to the absorbing state is 
related to a standard finite-size effect in the presence of absorbing states.
Any finite system settles in a stationary state with constant activity until
it hits the absorbing state because of a large spontaneous
fluctuation~\cite{Marrobook}.
The survival probability $P_s(t)$ that the system is still in an active
state  after time $t$ decays exponentially, $P_s(t)\sim\exp(-t/\tau)$.
Here $\tau$ is the  average lifetime in the active state and is found to
increase with the system size as $\tau \sim N$ (Fig.~\ref{Fig2}, inset).
This implies that in the thermodynamic limit ($N\to \infty$),
the system remains indefinitely in the stationary state, with everlasting
activity, i. e. incomplete ordering.
Hence, the voter model in the small-world regime behaves as its
mean field version (euclidean lattice with $d=\infty$)
that does not reach an ordered state~\cite{Rednerbook}.
This finding is quite interesting: in the thermodynamic limit,
the presence of long range connections does not make
the ordering process easier, rather it inhibits it.
The small-world topology of the network represents a barrier against
convergence to order. 

\begin{figure}
\onefigure[width=3.1in,angle=-90]{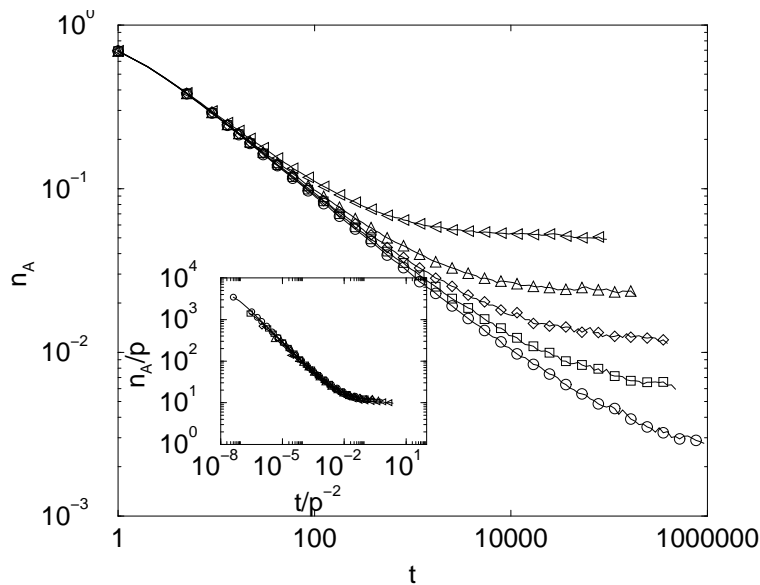}
\caption{Main: The fraction $n_A$ of active links as a function
of time for a large network with $N=10^5$ nodes. Data are for
$p=0.0002$ (circles), $p=0.0005$ (squares), $p=0.001$ (diamonds),
$p=0.002$ (triangles up) and $p=0.005$ (triangles left)
Inset: the same data, divided by $p$ and plotted as a function of $t/p^{-2}$
in order to show the validity of Eq.~(\ref{scaling}).}
\label{Fig3}
\end{figure}

In order to understand better the origin of  this incomplete ordering,
we study networks with $N=10^5$ nodes, for which the lifetime $\tau$ is much
larger than the time scales of interest and may therefore be considered
in practice as infinite.
In this case, as long as $p> 10^{-5}$ the network is in the small
world-regime and $n_A$ tends to a finite stationary value
(see Fig.~\ref{Fig3}). The network settles in a dynamically active
regime in which complete consensus does not emerge.
The value of $n_A$ in the stationary state depends on $p$.
In the one-dimensional lattice, the ordering process takes place via
free diffusion of domain boundaries (active bonds) and their annihilation
upon encounter.
In the small-world networks, shortcuts are an obstacle
for free diffusion of active bonds and tend to pin domain boundaries.
More in detail, the role played by sites connected with shortcuts
underlies the following
simple scaling analysis, valid for $t \gg 1$ and small $p$.
For short times domains form and start to coarsen:
$1/n_A$, their average size, is much smaller than the
average shortcut distance $\xi=2/(kp)$.
In this regime the evolution is practically equal to the
one-dimensional voter model, with $n_A$ decreasing as $t^{-1/2}$.
When the length of domains reaches the size of $\xi$, the only other length
in the problem, the behavior changes. 
The cross-over to the stationary state occurs therefore when $1/n_A\sim\xi$,
defining a diverging crossover time $t^* \sim p^{-2}$.
Since the crossover takes place at $n_A\sim p$ we obtain that 
\begin{equation} 
n_A(t,p) = p ~{\cal G}(t/p^{-2}),
\label{scaling}
\end{equation}
where the scaling function ${\cal G}(x)$ approaches a  constant value
for $x\gg1$.
Equation~(\ref{scaling}) is well obeyed by numerical results, that 
show a good data collapse on the predicted behavior (Fig.~\ref{Fig3}, inset). 

In summary we have shown that complex topological properties of 
small-world networks strongly affect the behavior of the voter model,
leading to incomplete ordering in the thermodynamical limit and to
counterintuitive phenomena for small systems.
We believe that the behavior of more general
models for social influence~\cite{Axelrod97b} is similarly modified
by the topology of the interaction network.
It would be interesting to test on these models
the effect of more heterogeneous topologies, such as scale-free networks,
which are known to alter several dynamical processes occurring on
them~\cite{Callaway00,Cohen01,Pastor00}.

During the completion of this work, we have become aware of some recent
work~\cite{Boyer02}
on the ordering process of the Ising model on the WS network,
presenting conclusions somewhat similar to ours.

\acknowledgments
We thank R. Pastor-Satorras for helpful comments and discussions.
This work has been partially supported by the European commission 
FET Open project COSIN IST-2001-33555.


\begin{thebibliography}{50}

\bibitem{Axelrod97}
\Name{Axelrod R.}
\Book{The complexity of cooperation}
\Publ{Princeton University Press, Princeton}
\Year{1997} 

\bibitem{Bray94}
\Name{Bray A. J.}
\REVIEW{Adv. Phys.}{43}{1994}{357}

\bibitem{Axtell96}
\Name{Axtell R., Axelrod R., Epstein J. \and Cohen M. D.}
\REVIEW{Computational and Mathematical Organization Theory}{1}{1996}{123}

\bibitem{Castellano00}
\Name{Castellano C., Marsili M. \and Vespignani A.}
\REVIEW{Phys. Rev. Lett.} {85} {2000} {3536}

\bibitem{Lopez01}
\Name{Sanchez A. D., Lopez J. M. \and Rodriguez M. A.}
\REVIEW{Phys. Rev. Lett.} {88} {2002}{048701}

\bibitem{Holyst01}
\Name{Holyst J. A., Kacperski K. \and Schweitzer F.}
\Book{Annual Reviews of Computational Physics}
\Vol{IX}
\Publ{World Scientific, Singapore}
\Year{2001}

\bibitem{Stauffer02}
\Name{Stauffer D.} 
\REVIEW{JASSS}{5}{2002}{1}

\bibitem{Liggett85}
\Name{Liggett T. M.}
\Book{Interacting Particle Systems}
\Publ{Springer, New York}
\Year{1985}

\bibitem{Vazquez02}
\Name{Vazquez F., Krapivsky P. L. \and Redner S.}
\Review{cond-mat/0209445}
\Year{2002}

\bibitem{note1}
In $d=1$ the voter model coincides with the Ising model with
Metropolis dynamics at $T=0$. In higher dimensions it differs 
from it because of the absence of surface tension;
\Name{Dornic I., Chat\'e H., Chave J. \and Hinrichsen H.}
\REVIEW{Phys. Rev. Lett.} {87} {2001} {045701}

\bibitem{Marrobook}
\Name{Marro, J. \and Dickman. R.}
\Book{Nonequilibrium Phase Transition in Lattice Models}
\Publ{Cambridge University Press, Cambridge}
\Year{1999}

\bibitem{Strogatz01}
\Name{Strogatz S. H.}
\REVIEW{Nature}{410}{2001}{268}

\bibitem{Barabasi02}
\Name{Albert R. \and Barab\'{a}si A.-L.}
\REVIEW{Rev. Mod. Phys.} {74} {2002}{47}

\bibitem{Watts98}
\Name{Watts D. J. \and Strogatz S. H.}
\REVIEW{Nature}{393}{1998}{440}

\bibitem{Wattsbook}
\Name{Watts D. J.}
\Book{Small Worlds: The Dynamics of Networks between Order and Randomness}
\Publ{Princeton University Press, Princeton}
\Year{1999}

\bibitem{Liljeros01}
\Name{Liljeros F., Edling C. R., Amaral L. A. N., Stanley H. E.
\and Aberg Y.}
\REVIEW{Nature} {411} {2001} {907}

\bibitem{Redner98}
\Name{Redner S.}
\REVIEW{Eur. Phys. J. B} {4} {1998}{131}

\bibitem{Newman00}
\Name{Newman M. E. J.}
\REVIEW{J. Stat. Phys.} {101} {2000}{819}

\bibitem{Barthelemy99}
\Name{Barthelemy M. \and Amaral L. A. N.}
\REVIEW{Phys. Rev. Lett.} {82} {1999} {3180}

\bibitem{Barrat00}
\Name{Barrat A. \and Weigt M.}
\REVIEW{Eur. Phys. J. B}{13}{2000}{547}

\bibitem{Rednerbook}
\Name{Frachebourg L. \and Krapivsky P. L.}
\REVIEW{Phys. Rev. E} {53} {1996}{3009};
\Name{Redner S.}
\Book{A Guide to First-Passage Processes}
\Publ{Cambridge University Press, Cambridge}
\Year{2001}

\bibitem{Axelrod97b}
\Name{Axelrod R.}
\REVIEW{J. of Conflict Resolut.} {41} {1997} {203}

\bibitem{Callaway00}
\Name{Callaway D. S., Newman M. E. J., Strogatz S. H. \and Watts D. J.}
\REVIEW{Phys. Rev. Lett.} {85} {2000} {5468}

\bibitem{Cohen01}
\Name{Cohen R., Erez K., ben-Avraham D. \and Havlin S.}
\REVIEW{Phys. Rev. Lett.} {85} {2000} {4626}

\bibitem{Pastor00}
\Name{Pastor-Satorras R. \and Vespignani A.}
\REVIEW{Phys. Rev. Lett.} {86} {2001} {3200}

\bibitem{Boyer02}
\Name{D. Boyer \and O. Miramontes}
\REVIEW{Phys. Rev. E} {67} {2003} {035102}


\end{thebibliography}
\end{document}